\documentclass[twocolumn,showpacs,amsfonts,aps,prl,amssymb, floatfix]{revtex4-1}
\usepackage{amsmath}
\usepackage{bm}
\usepackage{graphicx}
\usepackage{mathrsfs}
\usepackage{footmisc}
\usepackage{multirow}
\usepackage{graphicx}
\usepackage{booktabs, ctable}
\usepackage{xcolor, color, framed}

\begin{document}

\title{Internal Charmonium Evolution in the Quark-Gluon Plasma}


\author{Baoyi Chen$^1$, Xiaojian Du$^2$, Ralf Rapp$^2$}
\affiliation{
$^1$Department of Physics, Tianjin University, Tianjin 300350, China \\
$^2$Cyclotron Institute and Department of Physics and Astronomy, \\
Texas A\&M University, College Station, TX 77843-3366, USA}

\begin{abstract}
We employ a time-dependent Schr\"odinger equation to study the evolution of a $c\bar c$ dipole 
in a quark-gluon plasma (QGP). Medium effects on the heavy-quark potential in the QGP are found 
to significantly affect the timescales of the internal evolution of the dipole. Color-screening 
can enhance the overlap of the expanding wavepackage with excited states at high temperature, 
while it is reduced at lower temperatures where the dipole favors the formation of the charmonium 
ground state. We investigate the consequences of this mechanism on the double ratio of  
charmonium nuclear modification factors, $R_{AA}^{\psi(2S)}/R_{AA}^{J/\psi}$, 
in heavy-ion collisions. The impact of the transition mechanisms on this ratio turns out to be 
rather sensitive to the attractive strength of the potential, and to its temperature dependence. 
\end{abstract}
\pacs{25.75.-q, 12.38.Mh, 24.85.+p }
\maketitle





\section{I. Introduction}
The anomalous suppression of $J/\psi$ production in relativistic heavy-ion collisions
has been considered as a signal of the existence of deconfined matter, i.e., the 
``Quark-Gluon Plasma" (QGP)~\cite{Matsui:1986dk}.
In nucleus-nucleus collisions, the production of charmonia is affected by interactions 
within the incoming nuclei (cold nuclear matter, CNM) and the subsequently formed medium
(hot nuclear matter, HNM). The former include a shadowing of parton distribution functions, 
Cronin effect and nuclear absorption. They occur prior to the formation of QGP and,
at high energies, usually do not involve charmonium eigenstates, but rather pre-resonance 
states which therefore result in a similar modification of the eventually formed eigenstates.  
The HNM effects include inelastic collisions with partons, color screening and the 
recombination of charm and anti-charm quarks in QGP.  Differences between 
the measured nuclear modification factors ($R_{AA}$) of $J/\psi$ and $\psi(2S)$ should mainly 
be caused by the hot medium. The sequential suppression idea, based on the hierarchy of binding 
energies, suggests that $\psi(2S)$ should suffer much stronger suppression than $J/\psi$ in QGP, 
due to its smaller binding energy~\cite{Satz:2005hx}, with estimated dissociation temperatures 
of $T_d^{J/\psi}$$\sim$~1.5-2\,$T_c$ and $T_d^{\psi(2S)}$$\sim$$T_c$~\cite{Satz:2005hx}. 
This would mean that the $\psi(2S)$ can hardly survive in QGP and produce a small $R_{AA}^{\psi(2S)}$.
Indeed, applying an instantaneous melting picture with the above quoted temperature, transport model 
calculations result in a double ratio of $R_{AA}^{\psi(2S)}/R_{AA}^{J/\psi}$$\sim$0.1 in central 
Pb-Pb($\sqrt{s_{NN}}$=2.76\,TeV) collisions for 6.5\,GeV/c\,$<$\,$p_T$\,$<$\,30\,GeV/c~\cite{Chen:2013wmr}. 
This underestimates the experimental data significantly~\cite{Khachatryan:2014bva}. One idea to 
improve on this situation is to include regeneration processes for the $\psi'$, which, because 
of its smaller dissociation temperature, would likely be operative near the hadronization
transition and in the subsequent hadronic phase~\cite{Du:2015wha}.  
On the other hand, both QGP and the initially formed $Q\bar Q$ pairs in heavy-ion collisions 
have time evolution scales which can interfere and cause significant deviations from an 
equilibrium evolution~\cite{Kopeliovich:2015qna,Katz:2015qja}. This situation is reinforced
if the binding potential is weakened due to color screening, and for high-momentum charmonia
due to time dilation effects~\cite{Gerland:2003wy,Zhao:2008vu,Song:2015bja}.   
Instead of being an eigenstate, the $c\bar c$ dipole evolves as a superposition of eigenstates
through the medium. Most currently used quarkonium transport 
models~\cite{Zhao:2007hh,Zhu:2004nw,Strickland:2011aa,Ferreiro:2014bia,Song:2015bja,Chen:2015iga} 
do not account for transitions between ground and excited states. While this is probably
a good approximation for the predominantly produced ground states, a ``dynamical 
feeddown" may have significant effects on the yields of excited states~\cite{Sorge:1997bg}. 
The composition will not only change with time but also as a consequence
of medium effects on the potential, which specifically may enhance the overlap
with the larger-size excited states. 

\section{II. Time Dependent Schr\"odinger Equation}
Since the internal motion of $c$ and $\bar c$ inside charmonium bound states is 
approximately non-relativistic, we employ the time-dependent Schr\"odinger equation to 
describe its wavefunction evolution. We explore the in-medium 
heavy-quark (HQ) potential, $V(r,t)$,  within two limits, the free ($F(r,T)$) and internal 
$U(r,T)$ energies, for which we can employ results from lattice QCD~\cite{Digal:2005ht}. 
After separating off the angular parts, the radial time-dependent Schr\"odinger equation for the
quarkonium wavefunction can be written as
\begin{align}
\label{seq}
i\hbar {\partial \over \partial t}\psi( r, t) = [-{\hbar^2\over 2m_\mu}{\partial ^2\over \partial r^2} +V( r,t)]\psi(r,t)
\end{align}
where $r$ is the relative distance between $c$ and $\bar c$, $t$ is the proper time in
the center of mass frame, 
and $m_\mu=m_1m_2/(m_1+m_2)=m_c/2$ the reduced mass. 

To numerically solve Eq.~(\ref{seq}), we employ the Crank-Nicholson method~\cite{Crank:1996},
which utilizes a discretization of the time evolution as ${\bf T}^{n+1}\psi^{n+1} = \Gamma ^n$ with  
\begin{align}  
  \label{eq-1}
  \left(  
  \begin{array}{ccccc}  
          {\bf T}_{0,0}^{n+1} & {\bf T}_{0,1}^{n+1} & 0 &0 & \cdots \\
          {\bf T}_{1,0}^{n+1} & {\bf T}_{1,1}^{n+1}& {\bf T}_{1,2}^{n+1}& 0 & \cdots \\
          0 & {\bf T}_{2,1}^{n+1}& {\bf T}_{2,2}^{n+1}& {\bf T}_{2,3}^{n+1}& \cdots \\
          \cdots & \cdots & \cdots &\cdots &\cdots \\
          \cdots & \cdots & \cdots &\cdots &\cdots \\
  \end{array}  
  \right)  
  \left(
  \begin{array}{c}  
          \psi_{0}^{n+1} \\  
          \psi_{1}^{n+1}\\  
          \psi_{2}^{n+1}\\
          \psi_{3}^{n+1}\\
          \cdots  \\
 \end{array}  
 \right)
 &=  
  \left(
  \begin{array}{c}  
          \Gamma_{0}^{n} \\  
          \Gamma_{1}^{n}\\  
          \Gamma_{2}^{n}\\
          \Gamma_{3}^{n}\\
          \cdots  \\  
 \end{array} 
 \right) \ .
\end{align}
The non-zero entries are given by
\begin{align}
{\bf T}^{n+1}_{j,j}&= 2+2a+bV_j^{n+1}  \nonumber \\
{\bf T}^{n+1}_{j,j+1}&={\bf T}^{n+1}_{j+1,j}= -a \nonumber \\
\label{fun-CNsim}
\Gamma_j^n &= a\psi_{j-1}^n +(2-2a-bV_j^n)\psi_j^n +a\psi_{j+1}^n  
\end{align}
with $a= i\Delta t/(2m_\mu (\Delta r)^2)$ and $b=i\Delta t$.  

From the $c\bar c$ dipole wavefunction, $\psi(r,t)=rR( r,t)$,  
one can obtain the coefficient, $c_{mS}(t)$ ($m$=0,1,...), of 
a $S$-wave charmonium eigenstate in the $c\bar c$ dipole at each time step as 
\begin{align}
\label{cms-eq}
c_{mS}(t) &=  \int R_{mS}(r)e^{-iE_{mS} t} \psi(r,t)\cdot rdr \ . 
\end{align}
The pertinent nuclear modification factor of an eigenstate evolvig in a QGP fireball
can be defined as
\begin{align}
\label{RAA-eq}
R_{AA}^{mS}(t) &={\int d{\vec x}d\vec p |c_{mS}(t, \vec x, \vec p)|^2{dN_{c\bar c}\over d{\vec x}_0 d\vec p_{0}}{\mathcal{S}(\vec x_0,\vec p_{0})}\over 
\int d{\vec x}_0d\vec p_{0} |c_{mS}(0,{\vec x}_0, \vec p_{0})|^2{dN_{c\bar c}\over d{\vec x}_0 d\vec p_{0}}}
\end{align}
where $\vec x_0$ and $\vec p_0$ are the initial position and momentum of the center of 
$c\bar c$ dipole, with $\vec x= \vec x_0 +\vec v_{c\bar c}t$, $\vec v_{c\bar c}={\vec p_0}/E_{c\bar c}$ 
and  $\vec p =\vec p_0$ since we neglect in/elastic interactions of the $c\bar c$ dipole with the 
medium.
For the initial $c\bar c$ phase space distribution, $dN_{c\bar c}/d\vec x_0d\vec p_0$,
one usually assumes a nuclear collision profile, includes a shadowing factor,
and samples it using Monte-Carlo techniques.
However, we here constrain ourselves to evolve a $c\bar c$ dipole in a static medium, 
to focus on the color screening effects on the internal evolution of the wavepackage. 
The direct nuclear modification factor then simply becomes 
$R_{AA}^{mS}(t)=|c_{mS}(t)|^2/|c_{mS}(0)|^2$. 

\begin{figure}[!hbt]
\centering
\includegraphics[width=0.49\textwidth]{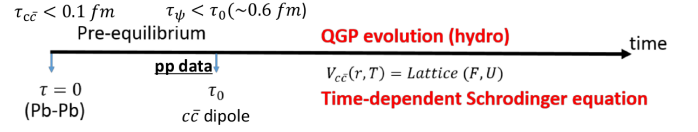}
\caption{Schematic diagram of a $c\bar c$ dipole evolution in heavy-ion collisions. 
After a hard collision in Pb-Pb collisions, the $c\bar c$ dipole is produced within 
$\tau_{c\bar c}<0.1$ fm/c. The vacuum formation time of charmonium eigenstates, $\tau_\psi$,
is usually comparable to the QGP thermalization time $\tau_0$, but can become significantly
longer at high momenta (Lorentz time dilation) and/or for color-screened potentials. 
}
\hspace{-0.1mm}
\label{fig-diagram}
\end{figure}
Finally, we need to specify our input for the initial $c\bar c$ dipole wavefunction. 
It can be a sharp Gaussian function during the production time, $\tau_{c\bar c}\lesssim0.1$\,fm/c.   
However, in $pp$ collisions, one should recover the observed fractions of charmonium states. 
This can be achieved by making a Gaussian ansatz,
\begin{align} 
\label{gauss} 
&\phi_0(r)=A \exp(-{r^2\over 2\sigma_0^2}) \ ,  
\end{align}
and adjust the the width, $\sigma_0$ accordingly. We find $\sigma_0$=0.23\,fm and 0.62\,fm to
satisfy the experimental value of $N_{pp}^{1S}/N_{pp}^{2S}$ in $pp$, representing 2 different
values of the relative velocity between $c$ and $\bar c$ inside the dipole. The coefficient  
$A=2/(\pi^{1/4}\sigma_0^{3/2)})$ can be fixed to the normalization of the radial wavefunction,
$\int r^2 dr \cdot |\phi_0(r)|^2 =1$. 
We neglect medium effects in the HQ potential in pre-equilibrium stage prior to QGP formation,  
which amounts to conserving the initial fraction of eigenstates in the wavepackage. 
Therefore, we start the $c\bar c$ evolution from $\tau_0$ where QGP reaches local equilibrium,
cf.~Fig.~\ref{fig-diagram}.

\section{III. Charmonium Evolution in Static Medium}
\begin{figure}[!hbt]
\centering
\includegraphics[width=0.4\textwidth]{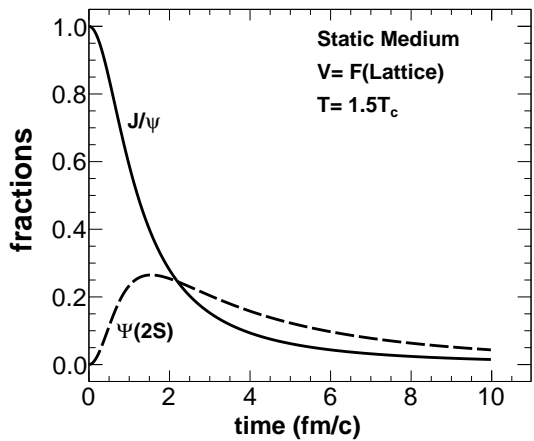}
\caption{The evolution of a stationary $c\bar c$ dipole in static medium with temperature 
$T= 1.5T_c$ and the free energy as the in-medium HQ potential. The initial wavefunction of 
the $c\bar c$ dipole is initialized as the $J/\psi$ eigenstate. Solid and dashed lines are 
the fractions of $J/\psi$ and $\psi(2S)$ in the $c\bar c$ wavefunction, respectively.
}
\hspace{-0.1mm}
\label{fig-jpsi-15Tc}
\end{figure}

We start with a stationary $c\bar c$ in a static QGP medium at $T$=1.5\,$T_c$, initialized 
as a pure $J/\psi$ state, cf.~fig.~\ref{fig-jpsi-15Tc}. With the HQ potential as the free energy, 
$F(r,T)$, the $J/\psi$ is a loosely bound dipole under these conditions. 
Its wavefunction expands outside, decreasing the overlap between the $c\bar c$ and the $J/\psi$ 
eigen-function, while increasing the overlap with the $\psi(2S)$ eigen-function for times 
$t$$<$2\,fm/$c$.  Subsequently, the $c\bar c$ dipole continues to expand decreasing the overlap 
with both $J/\psi$ and $\psi(2S)$ components, although the latter stays larger than the former.

\begin{figure}[!t]
\centering
\includegraphics[width=0.45\textwidth]{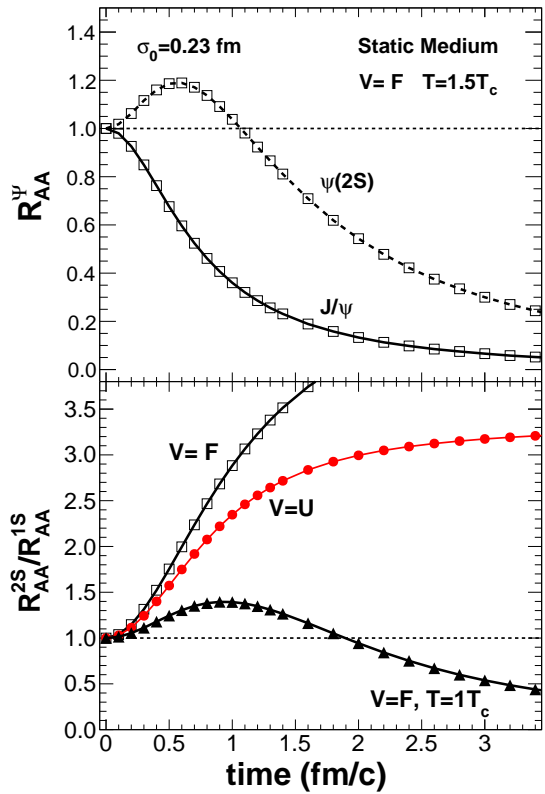}
\caption{(Color online)
Upper panel: Direct $J/\psi$ and $\psi(2S)$ nuclear modification factors, $R_{AA}$, as a 
function of time in a static medium with constant temperature, $T$=1.5\,$T_c$, and $V$=$F$.  
The initial $c\bar c$ wavefunction at $\tau_0$ is taken as a Gaussian function of width 
$\sigma_0$=0.23\,fm. Lower panel: double ratio, $R_{AA}^{2S}/R_{AA}^{1S}$, of direct $\psi'$ 
over $J/\psi$ corresponding to the curves in the upper panle (squares), and for 
$V$=$U$ (circles), as well as for a lower $T$=$T_c$ with $V$=$F$ (triangles). 
}
\hspace{-0.1mm}
\label{fig-sigma023}
\end{figure}

Next, we initialize the $c\bar c$ dipole wavefunction $\phi_0(r)$ at $\tau_0$ with 
the ratio of direct $J/\psi$ and $\psi(2S)$ yields in $pp$ collisions using Eq.~(\ref{gauss})
with $\sigma_0=0.23$\,fm, and evolve the $c\bar c$ dipole in the static medium with $T=1.5T_c$. 
For the HQ potential we consider both the free energy, $F(r,T)$, and internal energy, $U(r,T)$. 
Similar to the previous case, the $\psi(2S)$ fraction increases at first and then decreases 
with time, cf.~upper panel of Fig.~\ref{fig-sigma023}. The pertinent double ratio of direct 
$J/\psi$ and $\psi(2S)$ nuclear modification 
factors (without electromagnetic decay feeddowns from excited states), 
$R_{AA}^{2S}/R_{AA}^{1S}$, markedly increases with time, see lower panel of Fig.~\ref{fig-sigma023}. 
This effect is significantly less pronounced when using the internal energy as HQ potential, 
since the much stronger attraction tends to keep the $c\bar c$ wavepackage more compact thus
increasing (reducing) the overlap with the ground (excited) state, see the red line (with circles) 
in the lower panel of Fig.~\ref{fig-sigma023}.
A similar effect occurs at lower temperatures, e.g., at $T$=$T_c$. Even for the free energy,
a large overall reduction in the time dependence of the double ratio is found, dropping
below one for times $t$$>$2\,fm/c, cf.~the line with triangles in the lower panel of 
Fig.~\ref{fig-sigma023}.

\begin{figure}[!t]
\centering
\includegraphics[width=0.45\textwidth]{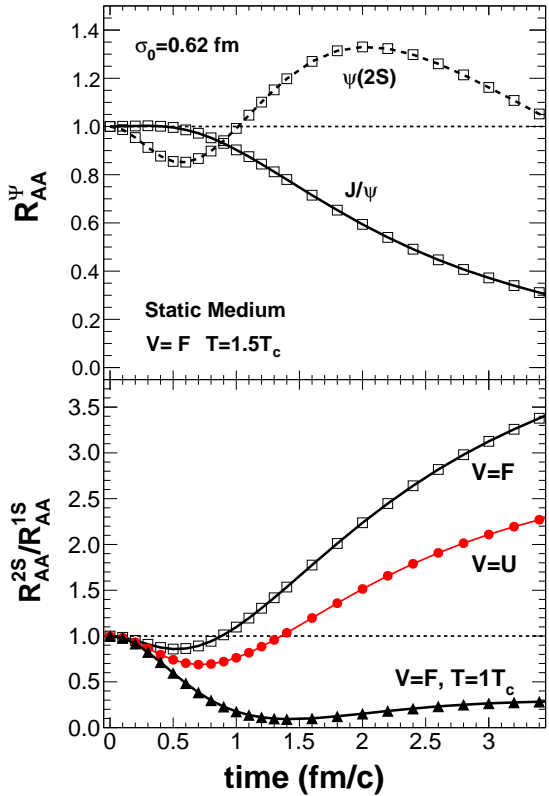}
\caption{(Color online)
Same as Fig.~\ref{fig-sigma023}, but for a larger Gaussian width of $\sigma_0$=0.62\,fm 
for the initial $c\bar c$ wavepackage at $\tau_0$. 
}
\hspace{-0.1mm}
\label{fig-sigma062}
\end{figure}

We repeat the same calculations as described in the previous paragraph for 
larger width of the initial Gaussian wavepackage, $\sigma_0=0.62$\,fm 
(which also recovers the empirical $\psi'$-over-$J/\psi$ ratio in $pp$ collisions),
cf.~Fig.\ref{fig-sigma062}. The enhancement in the $\psi'$ $R_{AA}$ is significantly
delayed, as is the suppression of the $J/\psi$ due to a smaller relative velocity
of $c$ and $\bar c$. These delays also manifest themselves in the slower
growth of the $R_{AA}^{2S}/R_{AA}^{1S}$ double ratio, starting out at values below
one for all 3 scenarios considered, while still reaching well above one at later times
for $T$=1.5\,$T_c$ and  both potential types (less for stronger attraction).
However, for temperatures near $T_c$, the double ratio drops dramatically within
the first 1\,fm/$c$ of the evolution and stays small thereafter. 

The inclusion of inelastic charmonium break-up reactions will introduce an imaginary
part into our calculation which will lead to a suppression of both $J/\psi$ and $\psi(2S)$,
which is most likely stronger for the excited states. Therefore, we expect the fractions 
in the lower panels of Figs.~\ref{fig-sigma023} and \ref{fig-sigma062} to be generally
reduced. However, the transition mechanisms, $J/\psi\leftrightarrow \psi(2S)$, still exist 
within the $c\bar c$ dipole and may well play an important role in the finally
observed $R_{AA}^{2S}/R_{AA}^{1S}$, including a nontrivial momentum dependence. 

We have also employed this approach to bottomonia. 
Due to the large binding energy of $\Upsilon(1S)$, the screening of the HQ potential 
at the relevant distances, $r\sim \langle r\rangle_{\Upsilon(1S)}$, is less pronounced
causing fewer transitions to occur. However, for the excited $\Upsilon(2S)$ and $\Upsilon(3S)$
states, we expect transitions to be active, and thus provide an independent handle from
the experimental side. This will be discussed in an upcoming manuscript. 

\section{IV. Conclusion}
In summary, we have employed the time-dependent Schr\"odinger equation to simulate the 
internal evolution of HQ dipoles in a hot medium. An in-medium potential causes
primordially produced correlated $c\bar c$ dipoles to undergo transitions between 
different asymptotic (vacuum) eigenstates over timescales which are significantly longer
than the vacuum formation times of these states. In particular, for weak binding the dipole
expands rather easily shifting wavefunction overlap from the ground to excited states. 
However, we have also found that whether these mechanisms result in an enhancement or 
suppression of the final double ratio, $R_{AA}^{2S}/R_{AA}^{1S}$, of the charmonium nuclear 
modification factors, rather sensitively depends on the strength of the binding potential.
A stronger attraction, i.e., less screening, due to either a stronger potential type (e.g., 
$U$ rather than $F$) or due to lower temperatures (e.g., near $T_c$), generally
leads to smaller double ratios. Related effects are expected to occur in the
$\Upsilon(2S)$-$\Upsilon(3S)$ sector.   
\\

\noindent
{\it Acknowledgment}. We thank P. Zhuang and Y. Liu for valuable discussions and input. 
This work has been supported by the US NSF under grant PHY-1614484, 
the NSFC and MOST grant Nos. 11335005, 
11575093, 2013CB922000, 2014CB845400 and 11547043.




%



\end{document}